\documentclass[twoside,11pt]{article}
\usepackage{jmlr2e_copy_11_6_18_giraffes}
\usepackage{subcaption}

\jmlrheading{volume TBD}{2018}{pages TBD}{other}{K. Gretchen Greene}{conference name}

\ShortHeadings{Expressive Augmentation and Complexity Layers}{Greene}
\begin{document}

\title{An Infinite Parade of Giraffes: Expressive Augmentation and Complexity Layers for Cartoon Drawing}

\author{\name K. G. Greene \email ggreene@media.mit.edu \\
	\addr MIT Media Lab\\
	Cambridge, MA, U.S.A.}

\maketitle

\begin{abstract}%
In this paper, we explore creative image generation constrained by small data. To partially automate the creation of cartoon sketches consistent with a specific designer's style, where acquiring a very large original image data set is impossible or cost prohibitive, we exploit domain specific knowledge for a huge reduction in original image requirements, creating an effectively infinite number of cartoon giraffes from just nine original drawings. We introduce ``expressive augmentations'' for cartoon sketches, mathematical transformations that create broad domain appropriate variation, far beyond the usual affine transformations, and we show that chained GANs models trained on the temporal stages of drawing or ``complexity layers'' can effectively add character appropriate details and finish new drawings in the designer's style. 

We discuss the application of these tools in design processes for textiles, graphics, architectural elements and interior design. 
\end{abstract}


\section{Introduction}

Computer vision deep learning tasks often require tens or hundreds of thousands of examples for training. Improved augmentation and training methods can enlarge and improve data sets and make new applications of deep learning possible where acquiring and labeling sufficient data was previously impossible or cost prohibitive. 

\begin{figure}
   \centering
   \begin{subfigure}[b]{0.08\linewidth}
      \includegraphics[width=\linewidth]{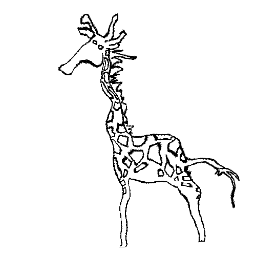}
   \end{subfigure}
   \begin{subfigure}[b]{0.08\linewidth}
      \includegraphics[width=\linewidth]{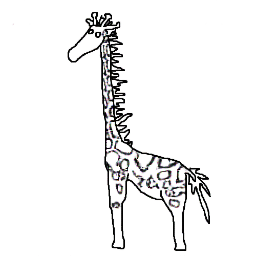}
   \end{subfigure}
   \begin{subfigure}[b]{0.08\linewidth}
      \includegraphics[width=\linewidth]{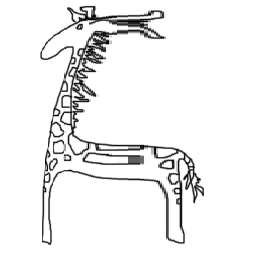}
   \end{subfigure}
   \begin{subfigure}[b]{0.08\linewidth}
      \includegraphics[width=\linewidth]{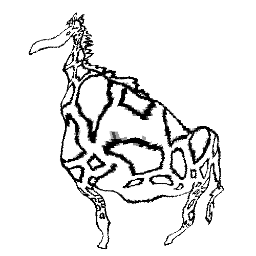}
   \end{subfigure}  
   \begin{subfigure}[b]{0.08\linewidth}
      \includegraphics[width=\linewidth]{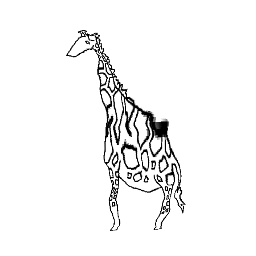}
   \end{subfigure}   
   \begin{subfigure}[b]{0.08\linewidth}
      \includegraphics[width=\linewidth]{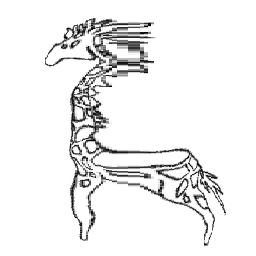}
   \end{subfigure}
   \begin{subfigure}[b]{0.08\linewidth}
      \includegraphics[width=\linewidth]{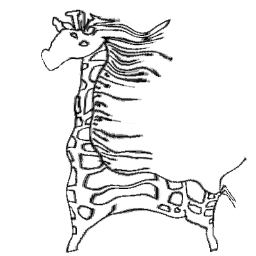}
   \end{subfigure}   
   \begin{subfigure}[b]{0.08\linewidth}
      \includegraphics[width=\linewidth]{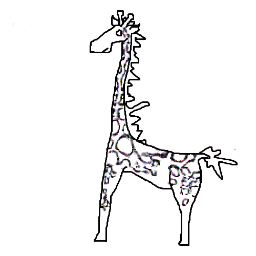}
   \end{subfigure}
   \begin{subfigure}[b]{0.08\linewidth}
      \includegraphics[width=\linewidth]{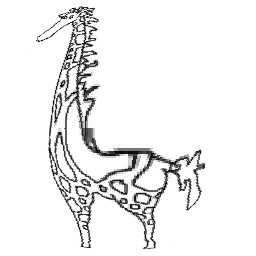}
   \end{subfigure}
   \caption{An infinite parade of giraffes...}
   \label{fig: parade}
\end{figure}

Using ``expressive augmentations'' to better represent drawing variations and ``complexity layers'', structuring the learning task as the temporal stages of drawing and chaining trained generative adversarial networks (GANs) models together, we partially automate the drawing of cartoon characters from extremely small original data sets, using just nine original drawings to create an effectively infinite parade of giraffes (Fig. \ref{fig: parade}). We show similar results with two other character types, flowers and dragons, creating pipelines that can create interesting families of variation from a single sketch and can add multiple character appropriate details to finish drawings: petals, manes, spots, spikes or wings, in a style consistent with the artist's.

Both the augmentations and the trained models become design tools, allowing for the rapid exploration of the character design space and the potential for ``infinite wallpaper'' and unique to the customer design variations at mass market prices. In addition, the careful study of process, augmentation and automation can lead to increased domain understanding and style expansion for the artist, the creator of the original cartoons.

\section{Deep Learning for Drawing and Coloring, Style Translation and Design Applications}

GANs, introduced by \citet{goodfellow:14}, have been used in a number of ways related to the automated generation and coloring of style consistent cartoon drawings. Applications include colorizing black and white landscapes, filling in missing sections of photographs and generating photographs of monster cats, shoes, handbags, building facades and street scenes from sketches or blocks of color \citep{pix2pix:16}, and turning horses into zebras and rendering photographs in the style of specific artists \citep{zhu:17}. Neural networks have been used to color Japanese cartoons \citep{paintschainer:17}, to generate simple cartoon sketches \citep{ha:17} and surrealist ``Deep Dream'' images \citep{mordvintsev:15}, and in website and graphic design production tools \citep{wired:17}. The smallest training data sets in these examples contained 400-1100 image pairs and most were much larger.

Relevant augmentation methods used on the MNIST hand written number data set include elastic deformations \citep{simard:03} and learned diffeomorphisms \citep{hauberg:16}.  

\section{Complexity Layers:  Learning the Temporal Stages of Drawings}

We started exploring extreme augmentation and GANs for drawing and design applications in \citet{greene:18}, where we combined GANs architecture borrowed from \citet{pix2pix:16} with geometric rules and elastic deformations to color cartoon flowers and dragons using original data sets of only 32-40 drawings. 

After successfully coloring cartoon characters from small data sets, we wondered whether the problem of drawing could be structured to use exactly the same GANs architecture and if our ideas for extreme augmentation could be expanded to allow us to work from even smaller, single digit original data sets. As in \citet{greene:18}, our post augmentation target would be 400-1100 AB training pairs of related 256 x 256 px images where A was the input and B was the desired output. Our original data goal would be 10-30 original drawings for each character type. 

The problem of generating totally new drawings from nothing doesn't look like the applications we'd seen in \citet{pix2pix:16}, but adding something to a drawing might. Maybe A could be a partially finished drawing and B the finished drawing, with the model acting like an artist's assistant, filling in some details in the artist's style after the artist draws the main form.

\begin{figure}
   \centering
   \begin{subfigure}[b]{0.08\linewidth}
      \includegraphics[width=\linewidth]{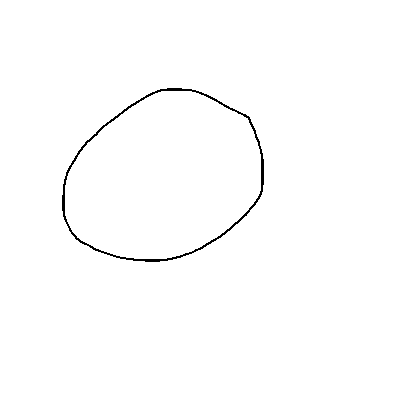}
   \end{subfigure}
   \begin{subfigure}[b]{0.08\linewidth}
      \includegraphics[width=\linewidth]{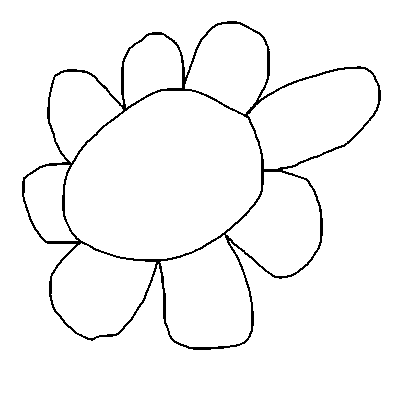}
   \end{subfigure}      
   \begin{subfigure}[b]{0.08\linewidth}
      \includegraphics[width=\linewidth]{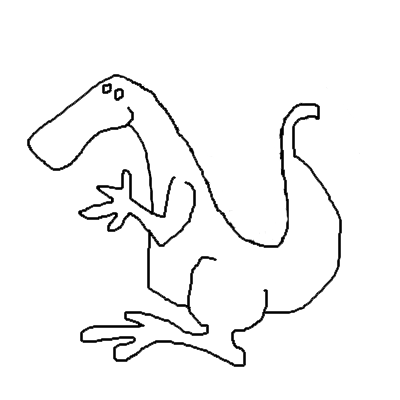}
   \end{subfigure}
   \begin{subfigure}[b]{0.08\linewidth}
      \includegraphics[width=\linewidth]{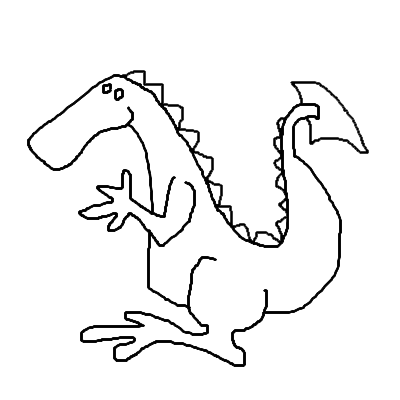}
   \end{subfigure}
   \begin{subfigure}[b]{0.08\linewidth}
      \includegraphics[width=\linewidth]{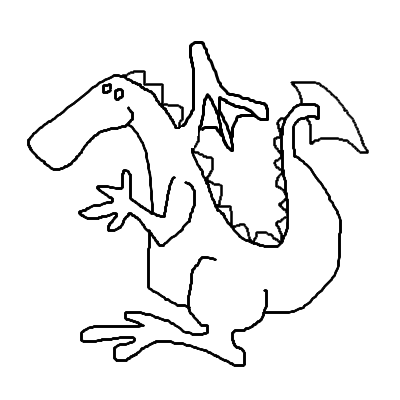}
   \end{subfigure}
   \begin{subfigure}[b]{0.08\linewidth}
      \includegraphics[width=\linewidth]{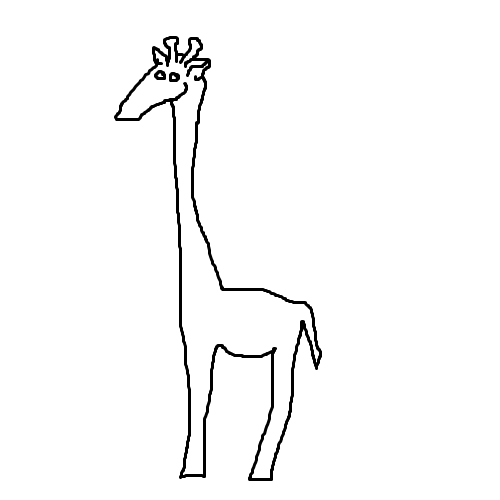}
   \end{subfigure}
   \begin{subfigure}[b]{0.08\linewidth}
      \includegraphics[width=\linewidth]{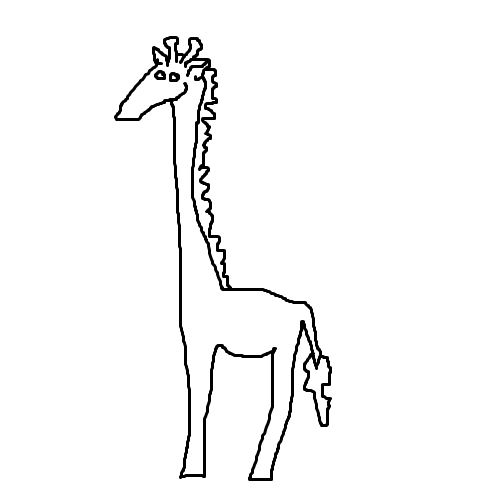}
   \end{subfigure}
   \begin{subfigure}[b]{0.08\linewidth}
      \includegraphics[width=\linewidth]{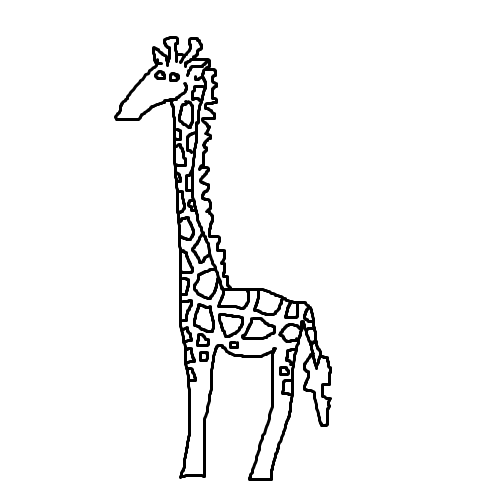}
   \end{subfigure}
   \caption{Temporal stages of drawing}
   \label{fig: ABC}
\end{figure}

We found that the creation of each of our character types could naturally be described as either two or three stages of adding a certain kind of detail, increasing complexity (Fig. \ref{fig: ABC}). Once a B or C drawing was finished, we couldn't efficiently go backwards, erasing details to create the simpler A image, but we could easily save intermediate drawings during the drawing process.

\textit{ }

\textit{Flower: }A = center, B = A + petals

\textit{Dragon: }A = body, B = A + spikes, C = B + wings

\textit{Giraffe: }A = body, B = A + mane, C = B + spots

\textit{ }

We hypothesized that exactly the same GANs model we'd used for coloring, trained instead on two temporal stages or complexity layers of drawing, would be able to add appropriate character details and also that we would get better performance if we trained two copies of the model, each on a single step, and chained them together sequentially, rather than trying to add all the complexity at once. For flowers, we would train the model on A to B and for dragons and giraffes, where there were three stages of drawing, we would train one copy of the model on A to B, another on B to C, and a third on A to C.

Before we could train our models and test our hypotheses, we needed to generate sufficiently large drawing data sets. With a target of 400-1100 AB or ABC training pairs or triples and a goal of reducing our original character drawing requirements below those used in our coloring work, we needed to continue to push the limits of extreme augmentation. 

\section{Beyond Affine: Expressive Augmentation for Cartoon Sketches}

Tall. Fat. Sneaky. Angry. Fancy. Smooth. Curvy. Bold. Shaky. The standard affine image augmentations: rotation, translation, flip and scale, do little to add the kind of variety to a drawn image data set that we would actually expect to see if an artist were to create additional originals for us. We would expect those new drawings to have expressive variation in drawn lines and character: a curvy line, a twisted tail, a happy grin, a pointy nose. Affine transformations can't create those kinds of variations, what we'll call ``expressive augmentations'', but there are other mathematical functions that might.

We explore two areas as sources for expressive augmentations: engineered homeomorphisms and elastic transformations and show the diversity of images we are able to create with them from a single original drawing. These expressive augmentations are intriguing image generation tools in their own right, allowing the artist to explore and expand their own style, adopting images as their own that they could not or would not have drawn directly. In addition, they provide enormous breadth in the synthetic data we can generate from each original drawing, and create the kinds of variation we are looking for, allowing us to increase our data set by orders of magnitude, successfully using GAN models trained on data sets created from as few as nine original drawings. 

\subsection{Engineered Homeomorphisms}

In our earlier work, \citet{greene:18}, we used a single engineered homeomorphism, $f_{daisy}(r,\theta)=(r^3,\theta)$ on the unit disc, to change sunflowers into daisies to augment our data set for coloring cartoon flowers. We build on that work here, expanding the kinds of homeomorphisms we use and the complexity of character types we use them on.

For engineered homeomorphisms, if we have a specific character type and a kind of geometric change that we're looking for, like shrinking a flower's center while stretching its petals or stretching a giraffe's neck without stretching its legs,  then we try to design a simple function in Cartesian or polar coordinates that makes that change. Our intuition for some kinds of transformations is better when we place images inside the unit disk, where it's easy to describe functions that keep the center and boundary fixed while stretching or twisting the rest. We also explored the effects of various functions without a specific effect in mind to develop our intuition and to discover interesting image transformations we hadn't thought to target.

\begin{figure}
   \centering
   \begin{subfigure}[b]{0.08\linewidth}
      \includegraphics[width=\linewidth]{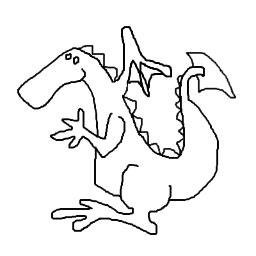}
   \end{subfigure}
      \begin{subfigure}[b]{0.08\linewidth}
      \includegraphics[width=\linewidth]{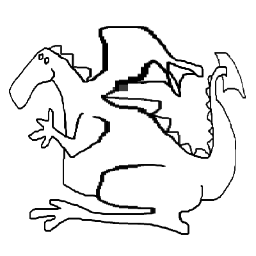}
   \end{subfigure}
   \begin{subfigure}[b]{0.08\linewidth}
      \includegraphics[width=\linewidth]{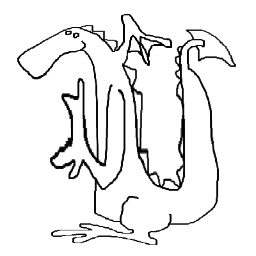}
   \end{subfigure}
   \begin{subfigure}[b]{0.08\linewidth}
      \includegraphics[width=\linewidth]{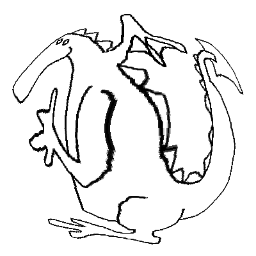}
   \end{subfigure}
   \begin{subfigure}[b]{0.08\linewidth}
      \includegraphics[width=\linewidth]{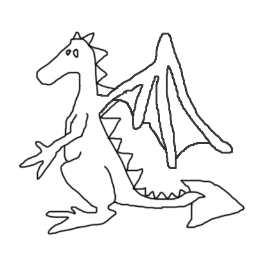}
   \end{subfigure}
      \begin{subfigure}[b]{0.08\linewidth}
      \includegraphics[width=\linewidth]{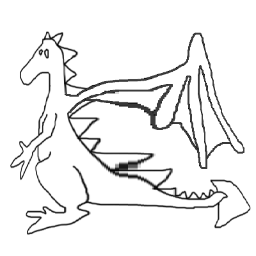}
   \end{subfigure}
   \begin{subfigure}[b]{0.08\linewidth}
      \includegraphics[width=\linewidth]{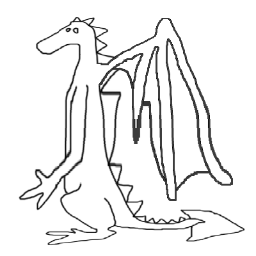}
   \end{subfigure}
   \begin{subfigure}[b]{0.08\linewidth}
      \includegraphics[width=\linewidth]{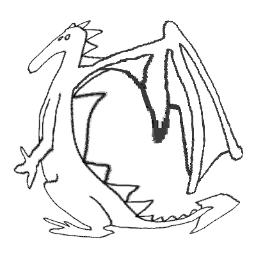}
   \end{subfigure}
   \caption{Two dragons: originals and homeomorphisms}
   \label{fig: dragon homeos}
\end{figure}

\begin{figure}
   \centering
   \begin{subfigure}[b]{0.08\linewidth}
      \includegraphics[width=\linewidth]{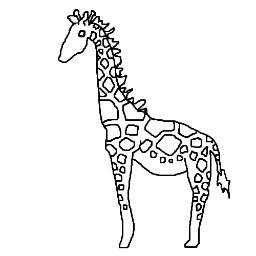}
   \end{subfigure}
   \begin{subfigure}[b]{0.08\linewidth}
      \includegraphics[width=\linewidth]{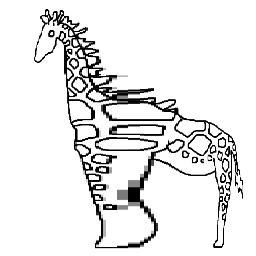}
   \end{subfigure}
   \begin{subfigure}[b]{0.08\linewidth}
      \includegraphics[width=\linewidth]{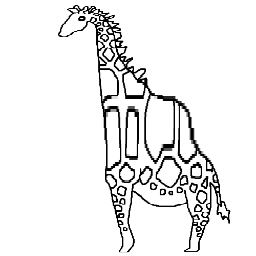}
   \end{subfigure}
   \begin{subfigure}[b]{0.08\linewidth}
      \includegraphics[width=\linewidth]{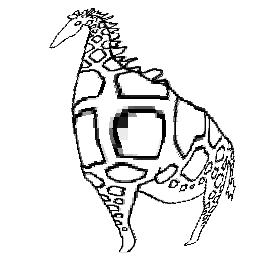}
   \end{subfigure}
   \begin{subfigure}[b]{0.08\linewidth}
      \includegraphics[width=\linewidth]{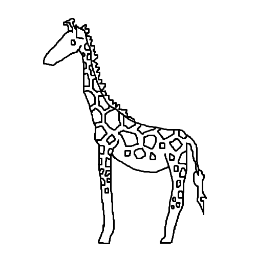}
   \end{subfigure}
   \begin{subfigure}[b]{0.08\linewidth}
      \includegraphics[width=\linewidth]{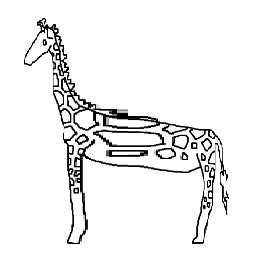}
   \end{subfigure}
   \begin{subfigure}[b]{0.08\linewidth}
      \includegraphics[width=\linewidth]{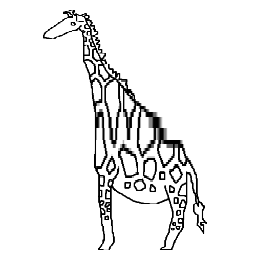}
   \end{subfigure}
   \begin{subfigure}[b]{0.08\linewidth}
      \includegraphics[width=\linewidth]{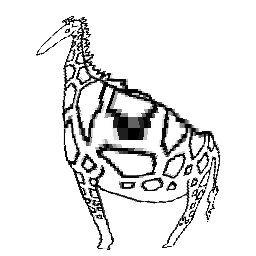}
   \end{subfigure}
   \caption{Two giraffes: originals and homeomorphisms}
   \label{fig: giraffe homeos}
\end{figure}   

For augmenting each of our character data sets, we used three homeomorphisms inside the unit disk, mapping our image into the unit disk and back out afterwards: $f_{xstretch}(x|x|,y) = (x,y)$, $f_{ystretch}(x,y|y|) = (x,y)$ and $f_{spherical}(x,y) = (x\sqrt{x^2+y^2},y)$ (Fig. \ref{fig: dragon homeos}). 

These transformations' effects are sensitive to the character's type and placement on the page. For example, $f_{xstretch}$ makes one giraffe's front leg unacceptably wide where the leg is at the center of the image, but makes a decent llama-giraffe on another giraffe where the middle of the image is in the body instead (Fig. \ref{fig: giraffe homeos}). This means a given transformation may work if and only if we reposition the character but it can also be used to increase the number of useful augmentations since small affine perturbations of a well placed character can provide a whole family of good variations.

\subsection{Elastic Deformation using Gaussian Filters}

Originally used by \citet{simard:03} on hand written numbers to augment the MNIST data set and proposed for broader document reading applications, we extended the use of elastic deformations to drawings in \citet{greene:18} to augment data sets for coloring cartoons and we use them here for creating the line drawings. In this paper, we substantially increase the number of elastic deformations applied to each original, amplifying their use in augmentation, and we also describe their usefulness beyond augmentation as an independent generative design tool. 

\begin{figure}
   \centering
   \begin{subfigure}[b]{0.08\linewidth}
      \includegraphics[width=\linewidth]{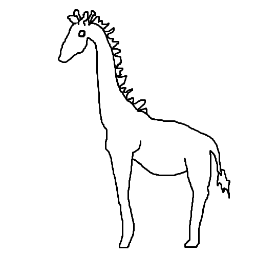}
   \end{subfigure}
   \begin{subfigure}[b]{0.08\linewidth}
      \includegraphics[width=\linewidth]{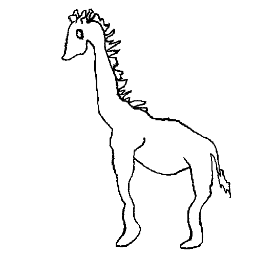}
   \end{subfigure}
   \begin{subfigure}[b]{0.08\linewidth}
      \includegraphics[width=\linewidth]{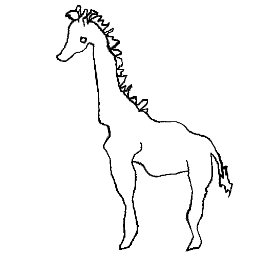}
   \end{subfigure}
   \begin{subfigure}[b]{0.08\linewidth}
      \includegraphics[width=\linewidth]{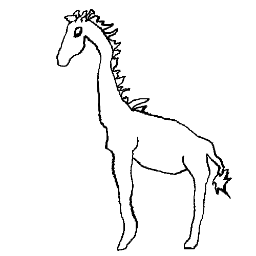}
   \end{subfigure}
   \begin{subfigure}[b]{0.08\linewidth}
      \includegraphics[width=\linewidth]{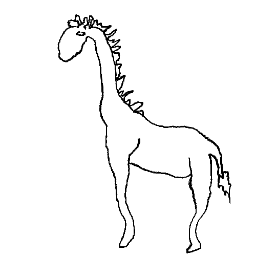}
   \end{subfigure}
   \begin{subfigure}[b]{0.08\linewidth}
      \includegraphics[width=\linewidth]{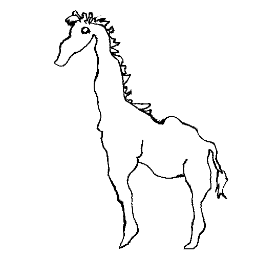}
   \end{subfigure}
   \begin{subfigure}[b]{0.08\linewidth}
      \includegraphics[width=\linewidth]{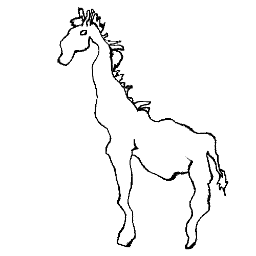}
   \end{subfigure}
   \caption{Some elastic deformations of one stage B giraffe}
   \label{fig: elastic giraffe}
\end{figure}

With the right parameter values, a random displacement field convolved with a Gaussian gives an elastic deformation that changes the character of the line and locally changes the drawing while leaving the overall pose of the cartoon character roughly the same. Starting from the same original giraffe drawing, an elastic deformation might make the nose bulbous or pointy, change the shape of the head, curl the tail, move the eye, or add some humps and bumps (Fig. \ref{fig: elastic giraffe}). 

\section{Experiments and Results}  

For three character types, flowers, dragons and giraffes, we divided the process of drawing into two (A and B) or three (A, B and C) stages as described in the complexity layers section. Drawings were made in a standard computer Paint program by the author and saved at intermediate and final stages. We augmented our original drawing pairs/triples with expressive augmentation and the usual affine transformations, using the same function or composition of functions with the same parameter values on each image in the pair or triple. We scaled our image size to 256x256 px.

We used Hesse's TensorFlow \citep{hesse:17} implementation of Isola et al.'s Pix2Pix image translation model \citep{pix2pix:16}, training copies of the model for 200 epochs on A to B, B to C or A to C. 

\subsection{Experiment 1: 30 Flowers - Centers to Petals}

\begin{figure}[h!]
   \centering
   \begin{subfigure}[b]{0.05\linewidth}
      \includegraphics[width=\linewidth]{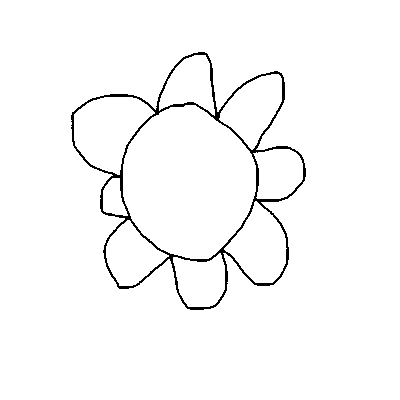}
   \end{subfigure}      
   \begin{subfigure}[b]{0.05\linewidth}
      \includegraphics[width=\linewidth]{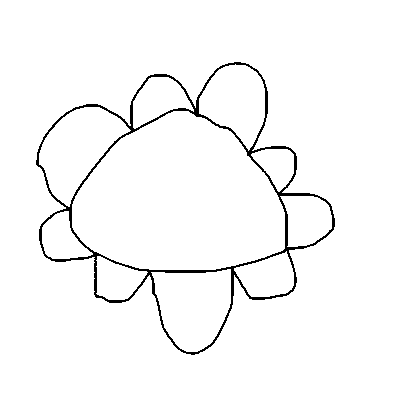}
   \end{subfigure}
   \begin{subfigure}[b]{0.05\linewidth}
      \includegraphics[width=\linewidth]{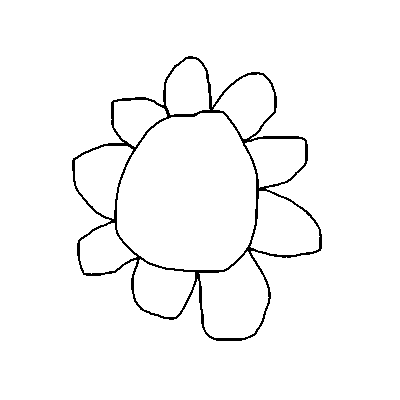}
   \end{subfigure}
   \begin{subfigure}[b]{0.05\linewidth}
      \includegraphics[width=\linewidth]{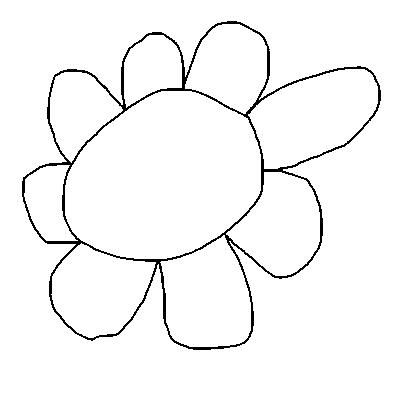}
   \end{subfigure}
   \begin{subfigure}[b]{0.05\linewidth}
      \includegraphics[width=\linewidth]{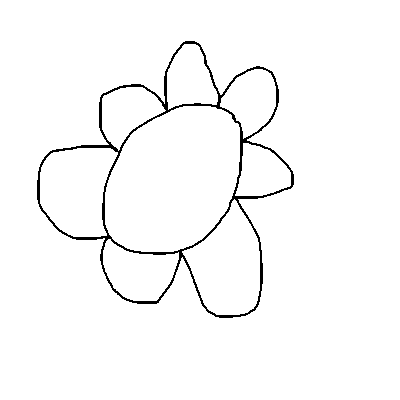}
   \end{subfigure}
   \begin{subfigure}[b]{0.05\linewidth}
      \includegraphics[width=\linewidth]{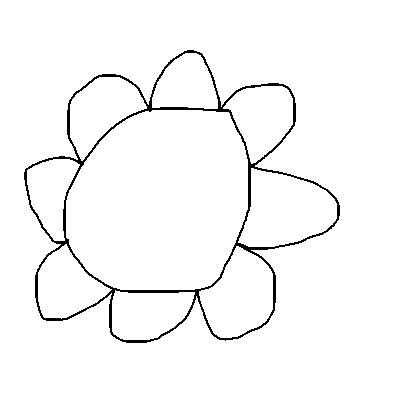}
   \end{subfigure}      
   \begin{subfigure}[b]{0.05\linewidth}
      \includegraphics[width=\linewidth]{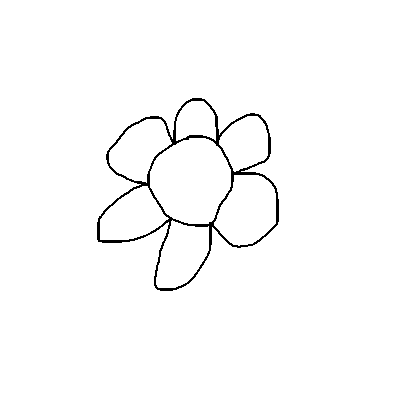}
   \end{subfigure}
   \begin{subfigure}[b]{0.05\linewidth}
      \includegraphics[width=\linewidth]{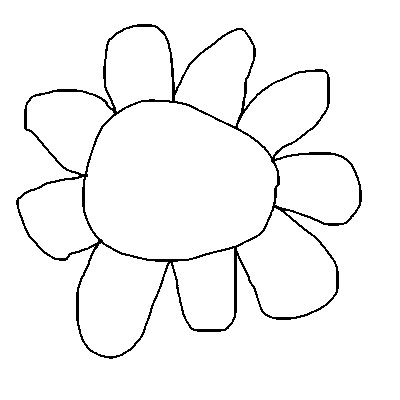}
   \end{subfigure}
   \begin{subfigure}[b]{0.05\linewidth}
      \includegraphics[width=\linewidth]{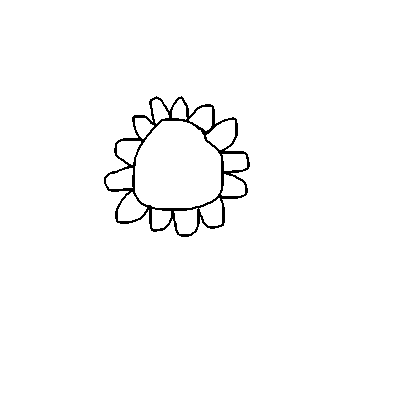}
   \end{subfigure}
   \begin{subfigure}[b]{0.05\linewidth}
      \includegraphics[width=\linewidth]{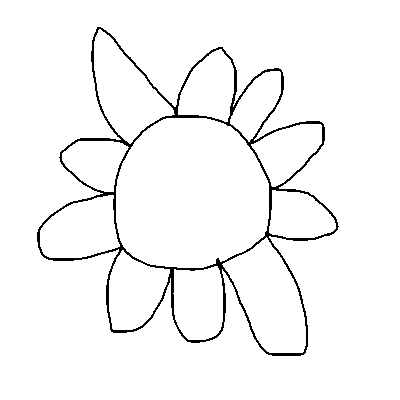}
   \end{subfigure}
   \begin{subfigure}[b]{0.05\linewidth}
      \includegraphics[width=\linewidth]{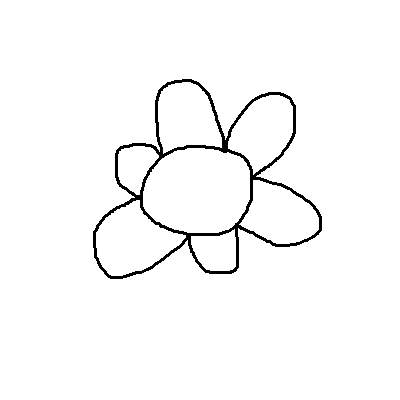}
   \end{subfigure}      
   \begin{subfigure}[b]{0.05\linewidth}
      \includegraphics[width=\linewidth]{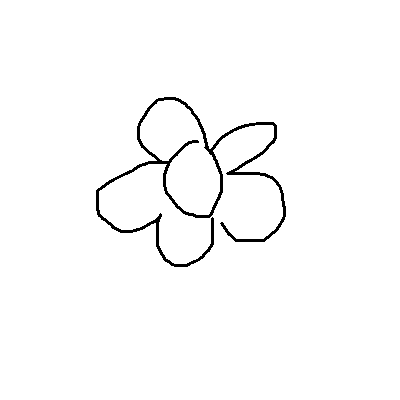}
   \end{subfigure}
   \begin{subfigure}[b]{0.05\linewidth}
      \includegraphics[width=\linewidth]{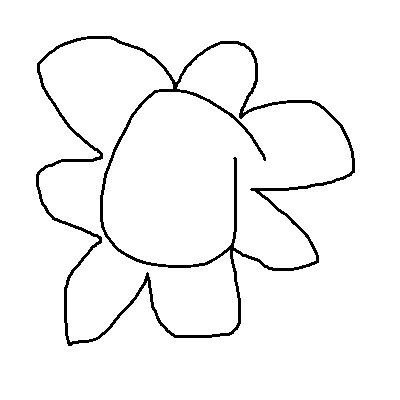}
   \end{subfigure}
   \begin{subfigure}[b]{0.05\linewidth}
      \includegraphics[width=\linewidth]{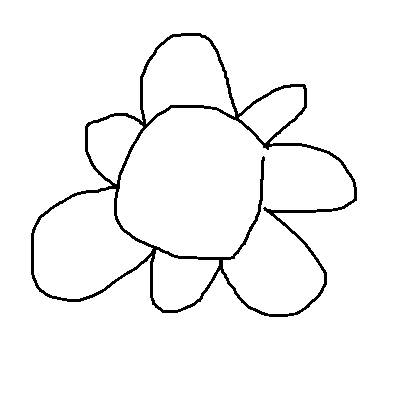}
   \end{subfigure}
   \begin{subfigure}[b]{0.05\linewidth}
      \includegraphics[width=\linewidth]{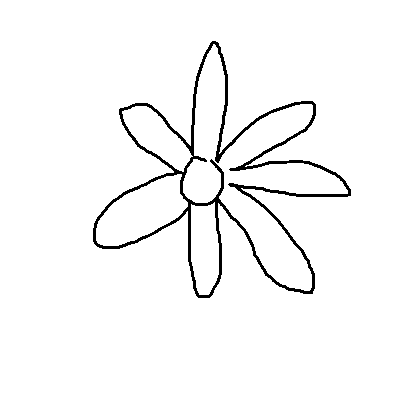}
   \end{subfigure}
   \begin{subfigure}[b]{0.05\linewidth}
      \includegraphics[width=\linewidth]{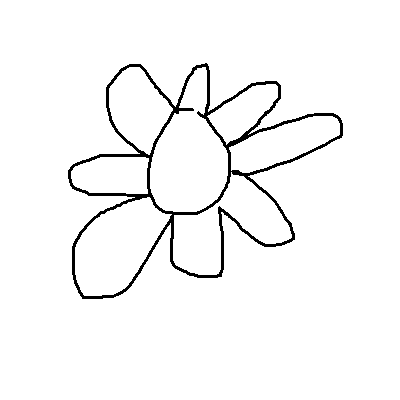}
   \end{subfigure}      
   \begin{subfigure}[b]{0.05\linewidth}
      \includegraphics[width=\linewidth]{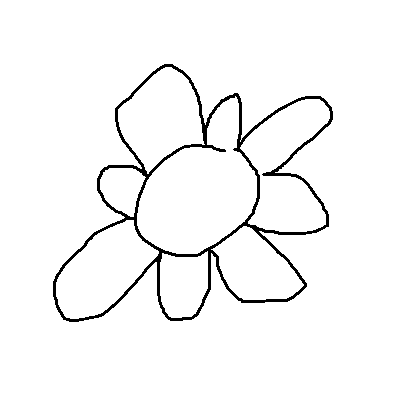}
   \end{subfigure}
   \begin{subfigure}[b]{0.05\linewidth}
      \includegraphics[width=\linewidth]{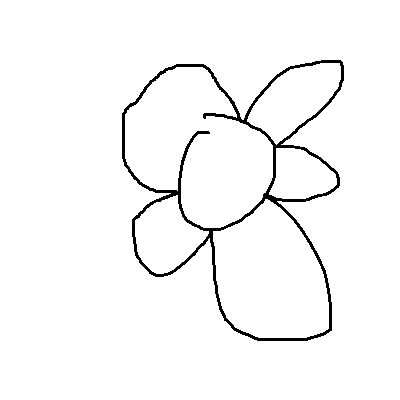}
   \end{subfigure}
   \begin{subfigure}[b]{0.05\linewidth}
      \includegraphics[width=\linewidth]{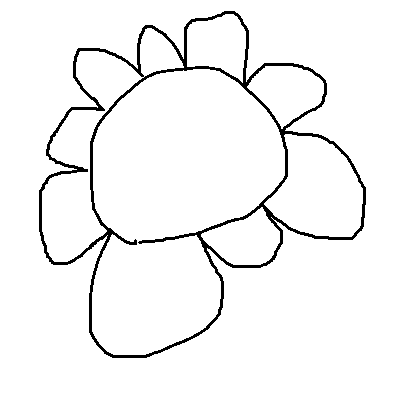}
   \end{subfigure}
   \begin{subfigure}[b]{0.05\linewidth}
      \includegraphics[width=\linewidth]{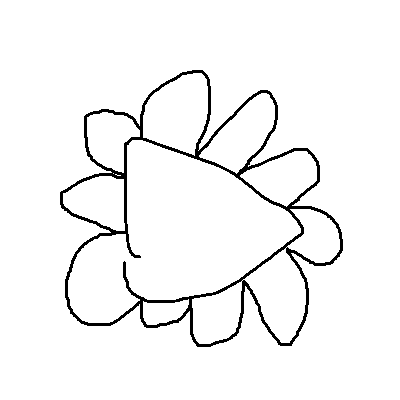}
   \end{subfigure}
   \begin{subfigure}[b]{0.05\linewidth}
      \includegraphics[width=\linewidth]{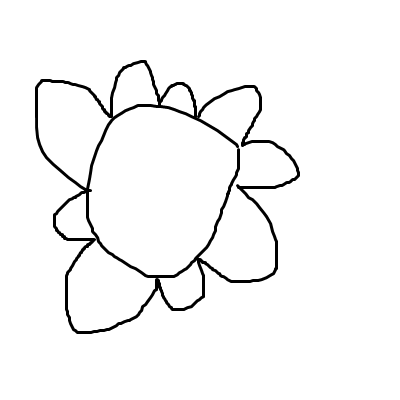}
   \end{subfigure}      
   \begin{subfigure}[b]{0.05\linewidth}
      \includegraphics[width=\linewidth]{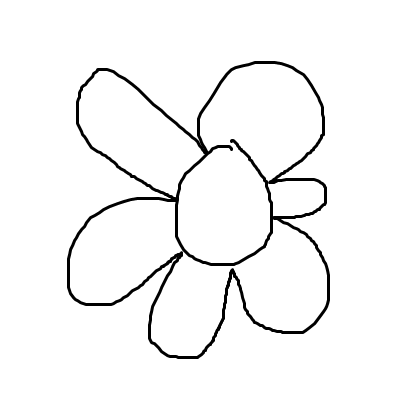}
   \end{subfigure}
   \begin{subfigure}[b]{0.05\linewidth}
      \includegraphics[width=\linewidth]{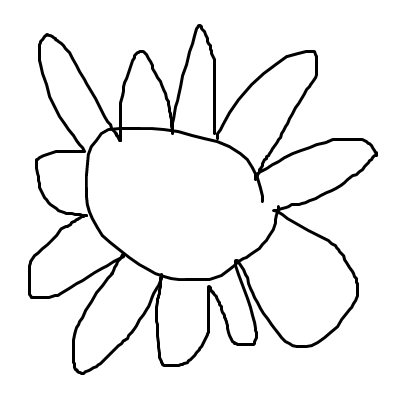}
   \end{subfigure}
   \begin{subfigure}[b]{0.05\linewidth}
      \includegraphics[width=\linewidth]{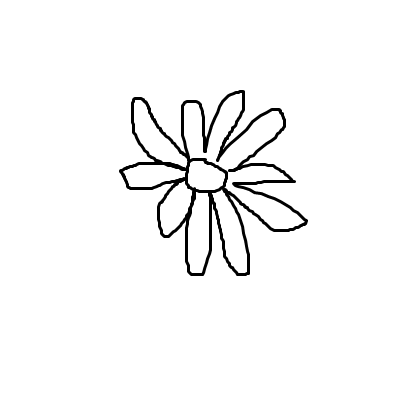}
   \end{subfigure}
   \begin{subfigure}[b]{0.05\linewidth}
      \includegraphics[width=\linewidth]{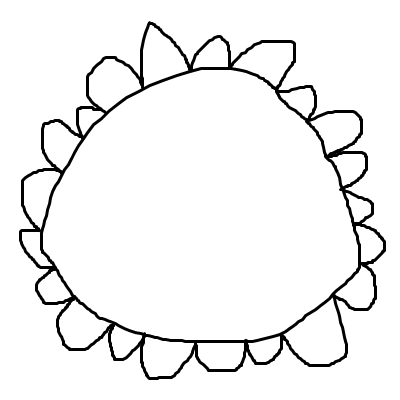}
   \end{subfigure}
   \begin{subfigure}[b]{0.05\linewidth}
      \includegraphics[width=\linewidth]{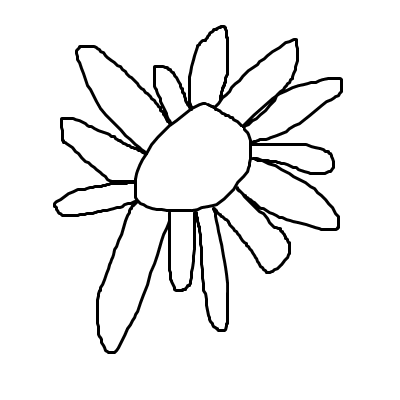}
   \end{subfigure}      
   \begin{subfigure}[b]{0.05\linewidth}
      \includegraphics[width=\linewidth]{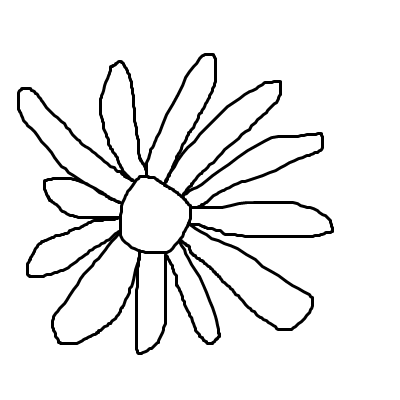}
   \end{subfigure}
   \begin{subfigure}[b]{0.05\linewidth}
      \includegraphics[width=\linewidth]{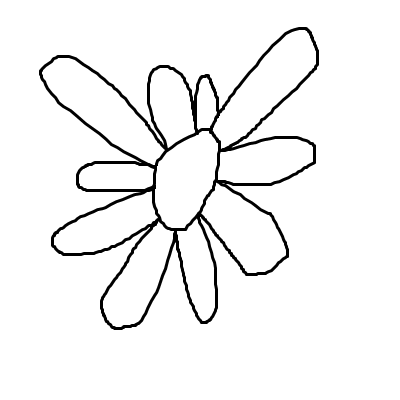}
   \end{subfigure}
   \begin{subfigure}[b]{0.05\linewidth}
      \includegraphics[width=\linewidth]{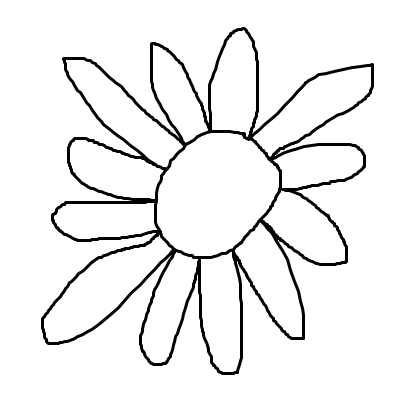}
   \end{subfigure}
   \begin{subfigure}[b]{0.05\linewidth}
      \includegraphics[width=\linewidth]{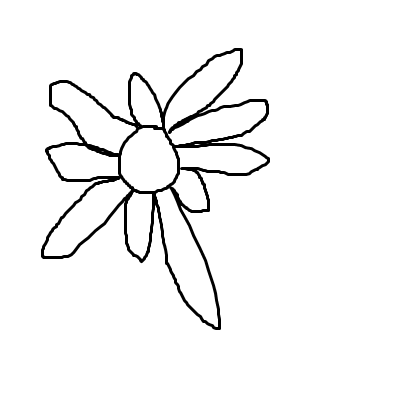}
   \end{subfigure}
   \caption{All 30 original stage B flower drawings}
   \label{fig: 30flowers}
\end{figure}

\textit{Training data: }Our set of 30 original flower drawings (Fig. \ref{fig: 30flowers}), saved in two stages, A (center) and B (center + petals) was augmented (with elastic deformations, affine transformations and our engineered homeomorphisms, $f_{xstretch}$, $f_{ystretch}$ and $f_{spherical}$) to create a training set of 645 A, B pairs. 

\begin{figure}
   \centering
   \begin{subfigure}[b]{0.05\linewidth}
      \includegraphics[width=\linewidth]{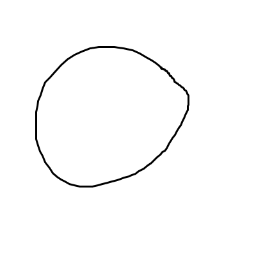}
   \end{subfigure}
   \begin{subfigure}[b]{0.05\linewidth}
      \includegraphics[width=\linewidth]{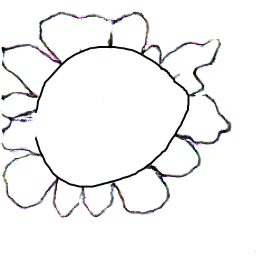}
   \end{subfigure}
   \begin{subfigure}[b]{0.05\linewidth}
      \includegraphics[width=\linewidth]{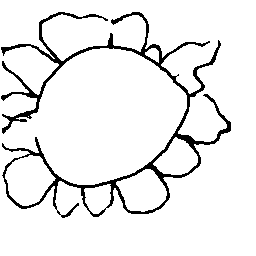}
   \end{subfigure}
   \begin{subfigure}[b]{0.05\linewidth}
      \includegraphics[width=\linewidth]{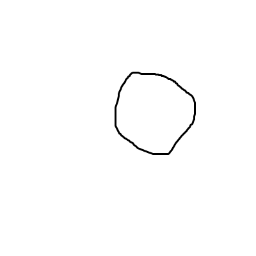}
   \end{subfigure}
   \begin{subfigure}[b]{0.05\linewidth}
      \includegraphics[width=\linewidth]{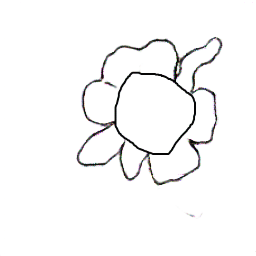}
   \end{subfigure} 
   \begin{subfigure}[b]{0.05\linewidth}
      \includegraphics[width=\linewidth]{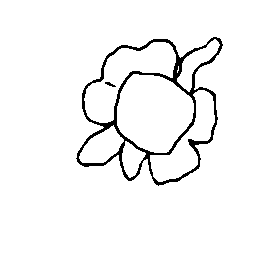}
   \end{subfigure}
   \begin{subfigure}[b]{0.05\linewidth}
      \includegraphics[width=\linewidth]{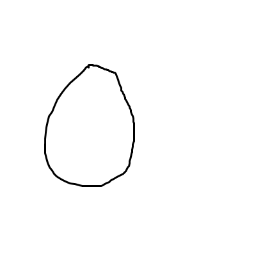}
   \end{subfigure}
   \begin{subfigure}[b]{0.05\linewidth}
      \includegraphics[width=\linewidth]{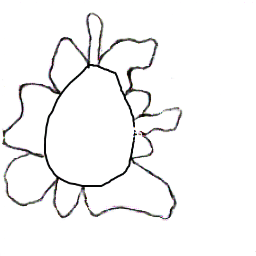}
   \end{subfigure}
   \begin{subfigure}[b]{0.05\linewidth}
      \includegraphics[width=\linewidth]{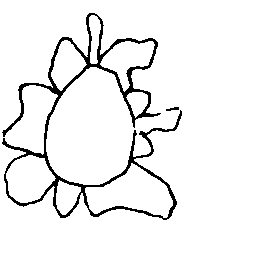}
   \end{subfigure}
   \begin{subfigure}[b]{0.05\linewidth}
      \includegraphics[width=\linewidth]{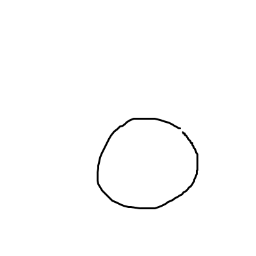}
   \end{subfigure}
   \begin{subfigure}[b]{0.05\linewidth}
      \includegraphics[width=\linewidth]{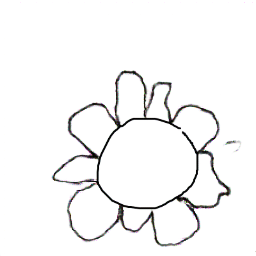}
   \end{subfigure}   
   \begin{subfigure}[b]{0.05\linewidth}
      \includegraphics[width=\linewidth]{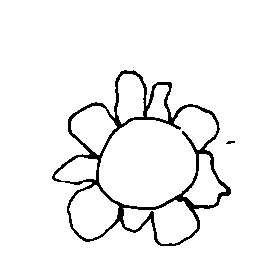}
   \end{subfigure}
   \begin{subfigure}[b]{0.05\linewidth}
      \includegraphics[width=\linewidth]{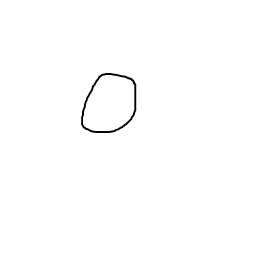}
   \end{subfigure}
   \begin{subfigure}[b]{0.05\linewidth}
      \includegraphics[width=\linewidth]{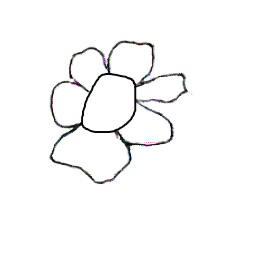}
   \end{subfigure}
   \begin{subfigure}[b]{0.05\linewidth}
      \includegraphics[width=\linewidth]{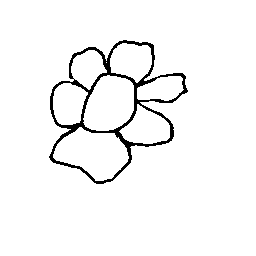}
   \end{subfigure}      
   \caption{Flowers: hand drawn centers, model drawn petals, Gaussian Otsu threshold}
   \label{fig: petals}
\end{figure}

\textit{Results: }Our trained AB model, given a new center, successfully draws that center with petals. The results show some narrow or poorly separated petals and the petal lines are not as dark and solid as the center lines, but overall the results are good. We got some improvement in the petal line quality postprocessing with Gaussian Otsu thresholding (Fig. \ref{fig: petals}). 

\subsection{Experiment 2: 20 Dragons - Bodies to Spikes to Wings}
\begin{figure}[h!]
   \centering
   \begin{subfigure}[b]{0.08\linewidth}
      \includegraphics[width=\linewidth]{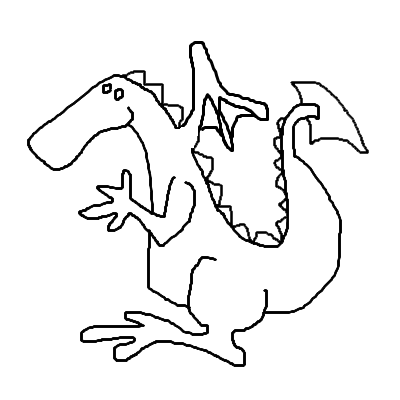}
   \end{subfigure}
   \begin{subfigure}[b]{0.08\linewidth}
      \includegraphics[width=\linewidth]{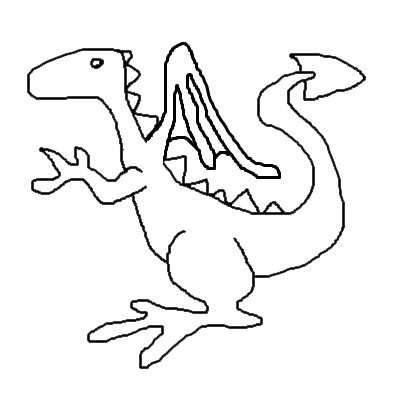}
   \end{subfigure}
   \begin{subfigure}[b]{0.08\linewidth}
      \includegraphics[width=\linewidth]{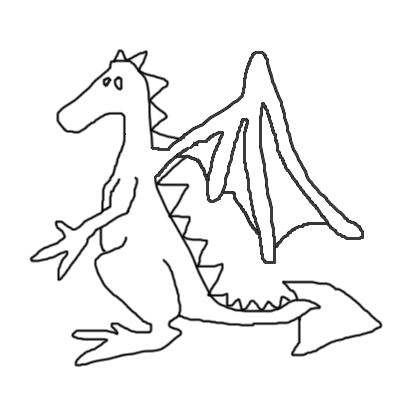}
   \end{subfigure}
   \begin{subfigure}[b]{0.08\linewidth}
      \includegraphics[width=\linewidth]{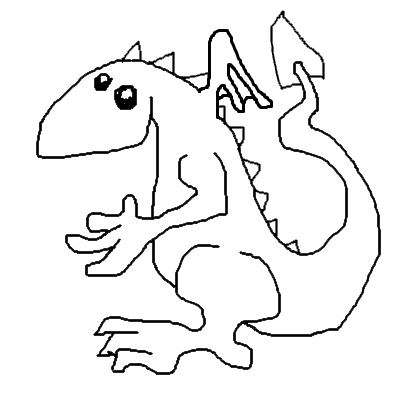}
   \end{subfigure}
   \begin{subfigure}[b]{0.08\linewidth}
      \includegraphics[width=\linewidth]{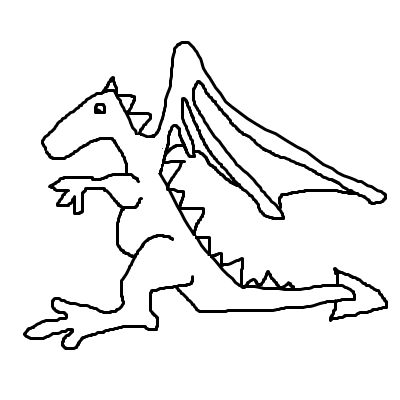}
   \end{subfigure}
   \begin{subfigure}[b]{0.08\linewidth}
      \includegraphics[width=\linewidth]{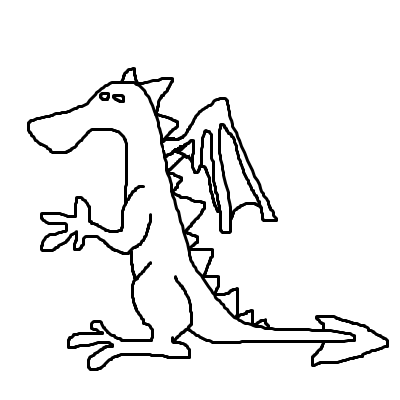}
   \end{subfigure}
   \begin{subfigure}[b]{0.08\linewidth}
      \includegraphics[width=\linewidth]{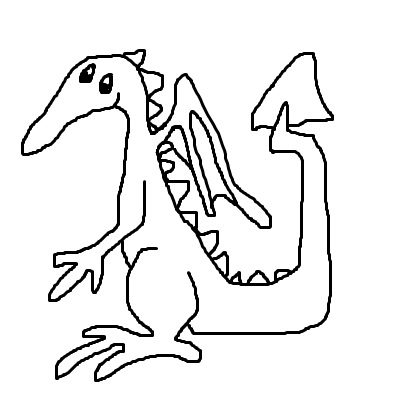}
   \end{subfigure}
   \begin{subfigure}[b]{0.08\linewidth}
      \includegraphics[width=\linewidth]{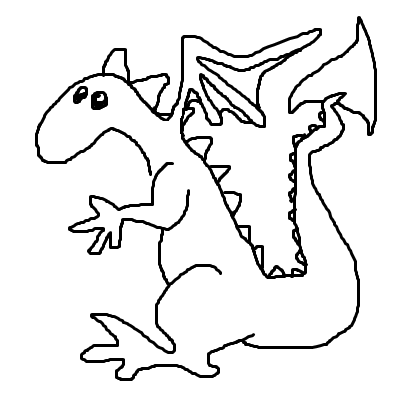}
   \end{subfigure}
   \begin{subfigure}[b]{0.08\linewidth}
      \includegraphics[width=\linewidth]{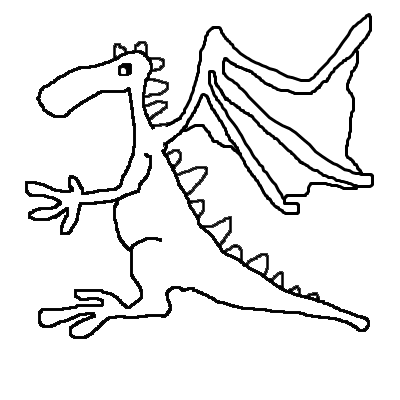}
   \end{subfigure}
   \begin{subfigure}[b]{0.08\linewidth}
      \includegraphics[width=\linewidth]{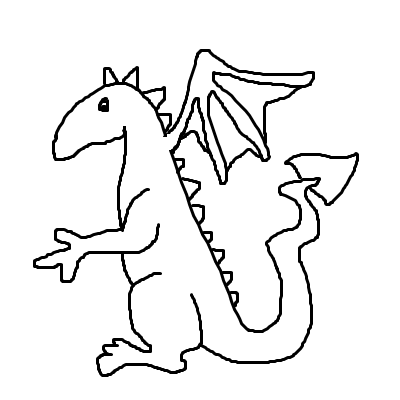}
   \end{subfigure}
   \begin{subfigure}[b]{0.08\linewidth}
      \includegraphics[width=\linewidth]{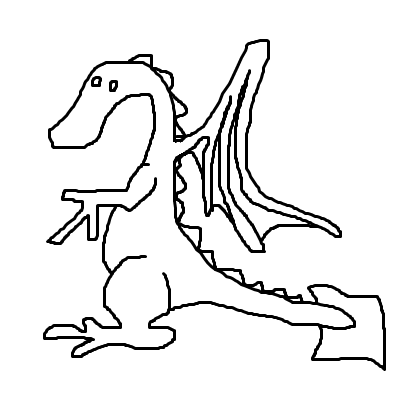}
   \end{subfigure}
   \begin{subfigure}[b]{0.08\linewidth}
      \includegraphics[width=\linewidth]{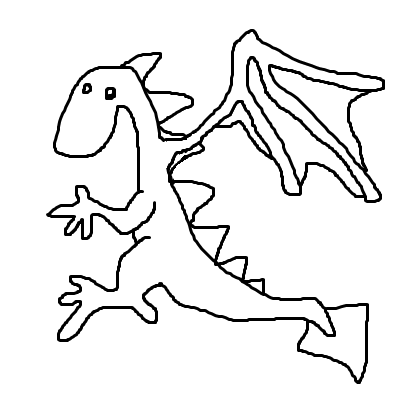}
   \end{subfigure}
   \begin{subfigure}[b]{0.08\linewidth}
      \includegraphics[width=\linewidth]{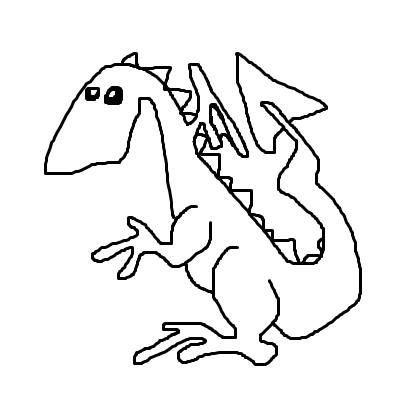}
   \end{subfigure}
   \begin{subfigure}[b]{0.08\linewidth}
      \includegraphics[width=\linewidth]{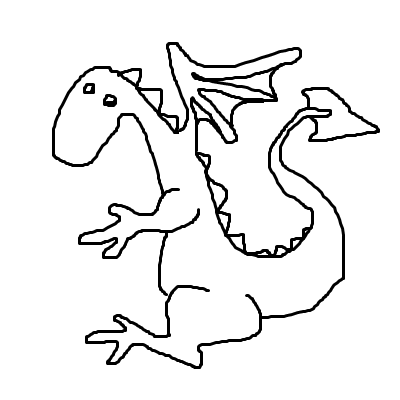}
   \end{subfigure}
   \begin{subfigure}[b]{0.08\linewidth}
      \includegraphics[width=\linewidth]{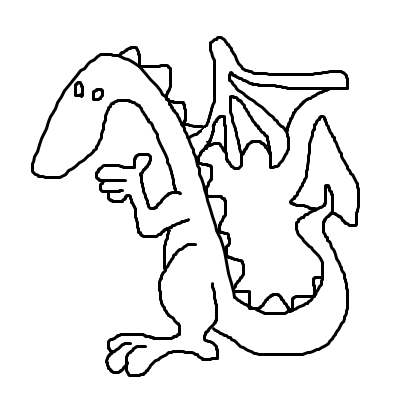}
   \end{subfigure}
   \begin{subfigure}[b]{0.08\linewidth}
      \includegraphics[width=\linewidth]{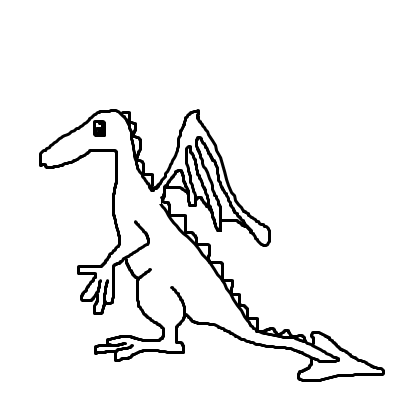}
   \end{subfigure}
   \begin{subfigure}[b]{0.08\linewidth}
      \includegraphics[width=\linewidth]{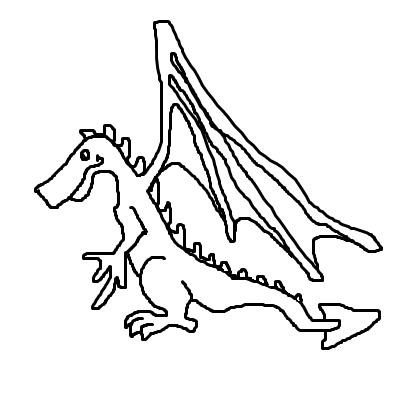}
   \end{subfigure}
   \begin{subfigure}[b]{0.08\linewidth}
      \includegraphics[width=\linewidth]{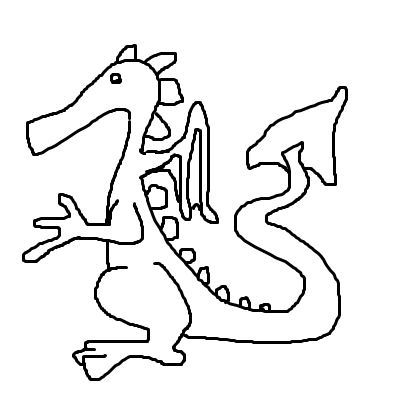}
   \end{subfigure}
   \begin{subfigure}[b]{0.08\linewidth}
      \includegraphics[width=\linewidth]{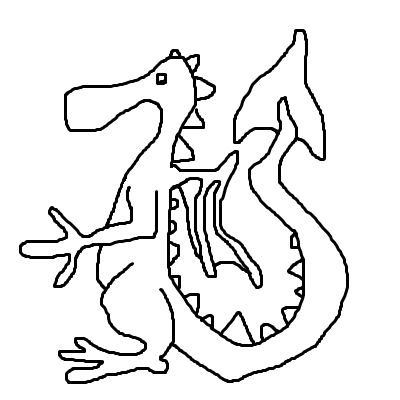}
   \end{subfigure}
   \begin{subfigure}[b]{0.08\linewidth}
      \includegraphics[width=\linewidth]{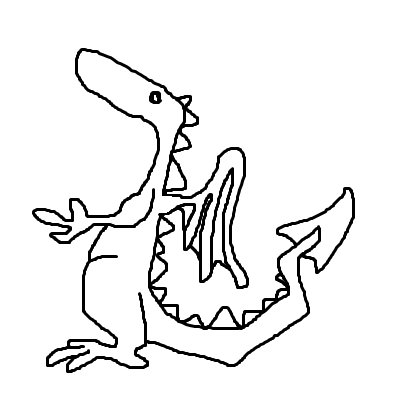}
   \end{subfigure}
   \caption{All 20 original stage C dragon drawings}
   \label{fig: 20dragons}
\end{figure}

\textit{Training data: }Our set of 20 original dragon drawings (Fig. \ref{fig: 20dragons}), saved in three stages, A (body), B (A + spikes), and C (B + wings) was augmented (with elastic deformations, affine and skew transformations and our engineered homeomorphisms, $f_{xstretch}$, $f_{ystretch}$ and $f_{spherical}$) to create a training set of 645 A, B, C triples. 

\begin{figure}
   \centering
   \begin{subfigure}[b]{0.08\linewidth}
      \includegraphics[width=\linewidth]{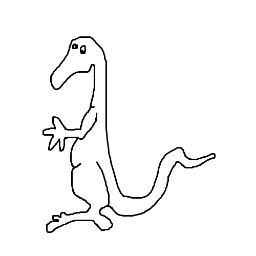}
   \end{subfigure}      
   \begin{subfigure}[b]{0.08\linewidth}
      \includegraphics[width=\linewidth]{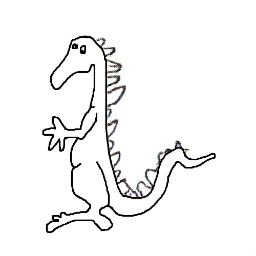}
   \end{subfigure} 
   \begin{subfigure}[b]{0.08\linewidth}
      \includegraphics[width=\linewidth]{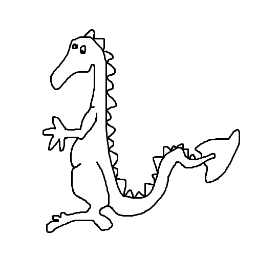}
   \end{subfigure}   
   \begin{subfigure}[b]{0.08\linewidth}
      \includegraphics[width=\linewidth]{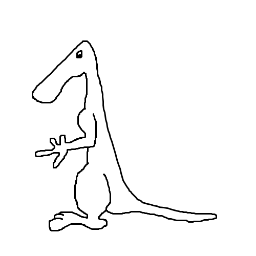}
   \end{subfigure}      
   \begin{subfigure}[b]{0.08\linewidth}
      \includegraphics[width=\linewidth]{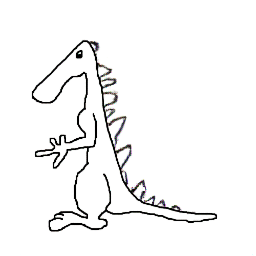}
   \end{subfigure} 
   \begin{subfigure}[b]{0.08\linewidth}
      \includegraphics[width=\linewidth]{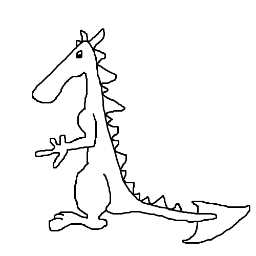}
   \end{subfigure} 
   \caption{Dragon spikes: hand drawn A, AB model drawn, hand drawn target B}
   \label{fig: ABspikes}
\end{figure}

\begin{figure}
   \centering      
   \begin{subfigure}[b]{0.08\linewidth}
      \includegraphics[width=\linewidth]{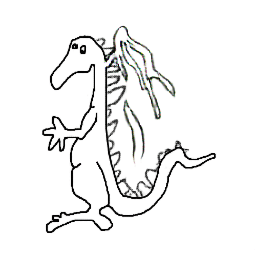}
   \end{subfigure} 
   \begin{subfigure}[b]{0.08\linewidth}
      \includegraphics[width=\linewidth]{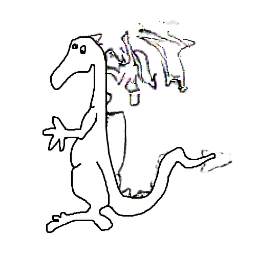}
   \end{subfigure} 
   \begin{subfigure}[b]{0.08\linewidth}
      \includegraphics[width=\linewidth]{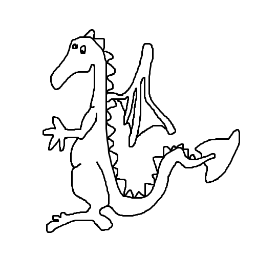}
   \end{subfigure}         
   \begin{subfigure}[b]{0.08\linewidth}
      \includegraphics[width=\linewidth]{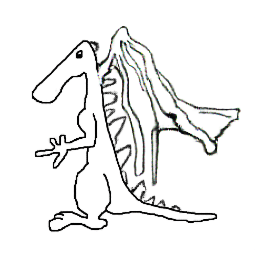}
   \end{subfigure} 
   \begin{subfigure}[b]{0.08\linewidth}
      \includegraphics[width=\linewidth]{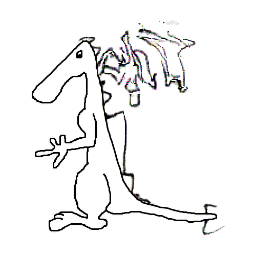}
   \end{subfigure} 
   \begin{subfigure}[b]{0.08\linewidth}
      \includegraphics[width=\linewidth]{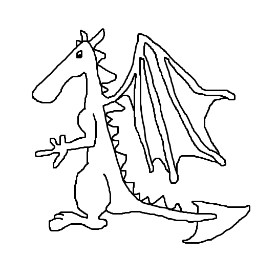}
   \end{subfigure} 
   \caption{Dragon chain test: AB + BC models, AC model, hand drawn target C}
   \label{fig: dragonchaintest}
\end{figure}

\textit{Results: }Our AB trained model, given a new body, draws that body with good back spikes but no tail spike (Fig. \ref{fig: ABspikes}). Our BC trained model, given a new body with spikes adds fair wings. If we chain the AB and BC models together, from a new body we get good back spikes and fair wings. Our AC trained model, which tries to learn everything in one step, is much worse on wings and arguably worse on both features (Fig. \ref{fig: dragonchaintest}).  

\subsection{Experiment 3: 9 Giraffes - Bodies to Manes to Spots}
\begin{figure}[h!]
   \centering
   \begin{subfigure}[b]{0.08\linewidth}
      \includegraphics[width=\linewidth]{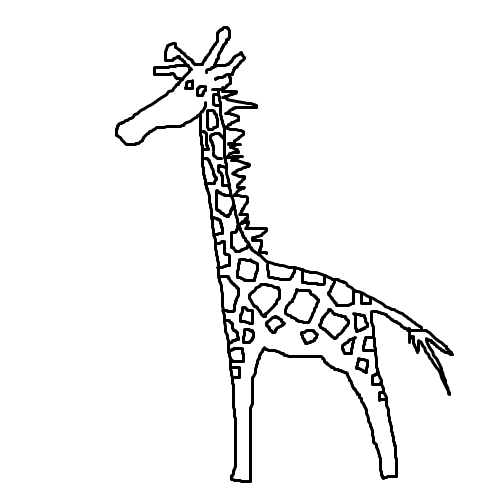}
   \end{subfigure}
   \begin{subfigure}[b]{0.08\linewidth}
      \includegraphics[width=\linewidth]{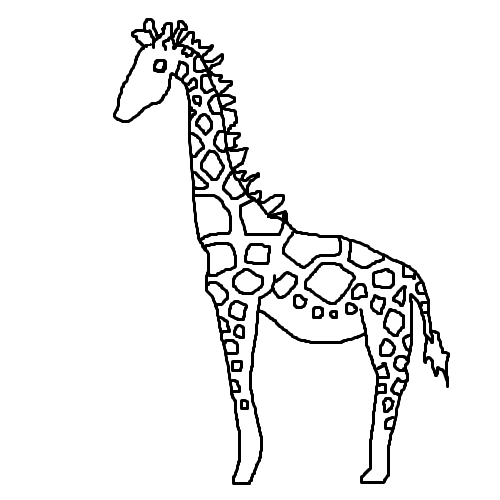}
   \end{subfigure}
   \begin{subfigure}[b]{0.08\linewidth}
      \includegraphics[width=\linewidth]{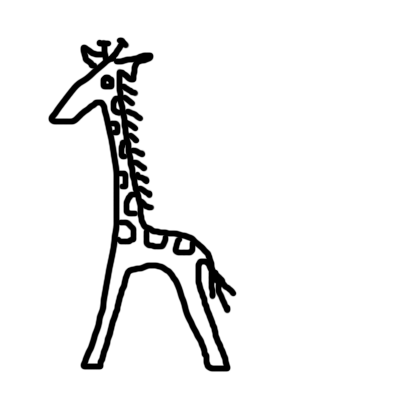}
   \end{subfigure}
   \begin{subfigure}[b]{0.08\linewidth}
      \includegraphics[width=\linewidth]{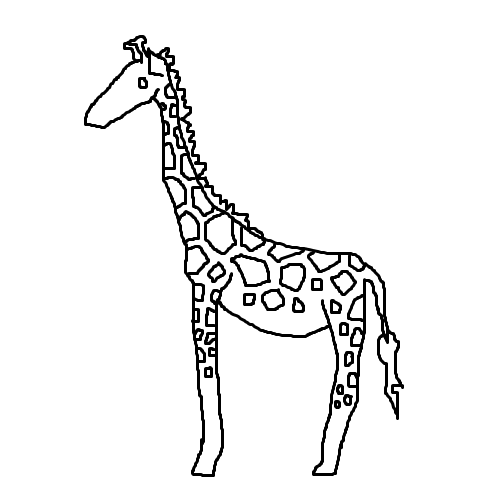}
   \end{subfigure}
   \begin{subfigure}[b]{0.08\linewidth}
      \includegraphics[width=\linewidth]{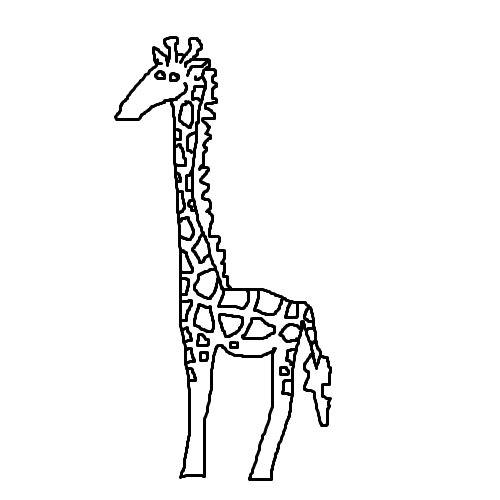}
   \end{subfigure}
   \begin{subfigure}[b]{0.08\linewidth}
      \includegraphics[width=\linewidth]{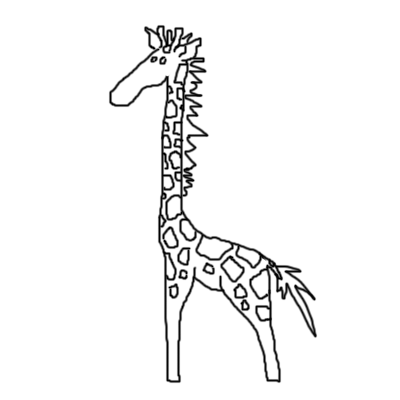}
   \end{subfigure}
   \begin{subfigure}[b]{0.08\linewidth}
      \includegraphics[width=\linewidth]{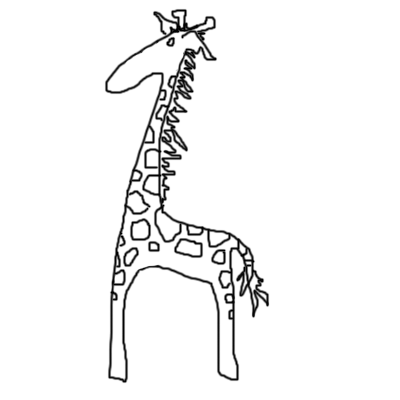}
   \end{subfigure}
   \begin{subfigure}[b]{0.08\linewidth}
      \includegraphics[width=\linewidth]{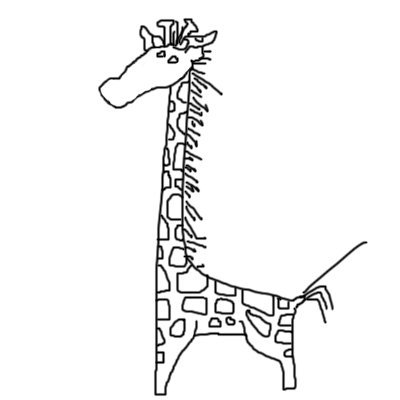}
   \end{subfigure}
   \begin{subfigure}[b]{0.08\linewidth}
      \includegraphics[width=\linewidth]{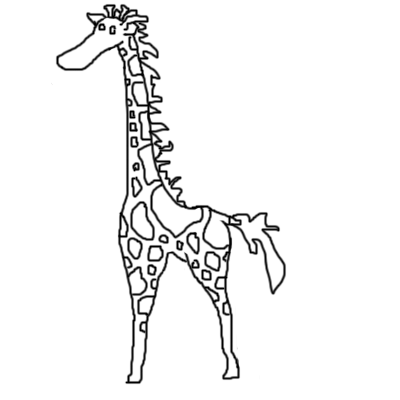}
   \end{subfigure}
   \caption{All nine original stage C giraffe drawings}
   \label{fig: 9giraffes}
\end{figure}

\textit{Training data: }Our set of 9 original giraffe drawings (\ref{fig: 9giraffes}), saved in three stages, A (body), B (A + mane and tail), and C (B + spots) was augmented (with elastic deformations, affine transformations and our engineered homeomorphisms, $f_{xstretch}$, $f_{ystretch}$ and $f_{spherical}$) to create a training set of 398 A, B, C triples. 

\begin{figure}
   \centering
   \begin{subfigure}[b]{0.08\linewidth}
      \includegraphics[width=\linewidth]{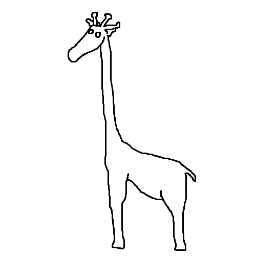}
   \end{subfigure}
   \begin{subfigure}[b]{0.08\linewidth}
      \includegraphics[width=\linewidth]{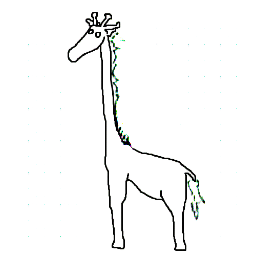}
   \end{subfigure}
   \begin{subfigure}[b]{0.08\linewidth}
      \includegraphics[width=\linewidth]{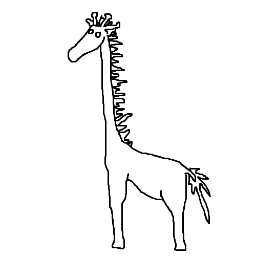}
   \end{subfigure}
   \begin{subfigure}[b]{0.08\linewidth}
      \includegraphics[width=\linewidth]{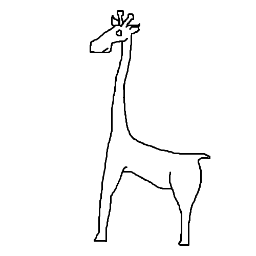}
   \end{subfigure}
   \begin{subfigure}[b]{0.08\linewidth}
      \includegraphics[width=\linewidth]{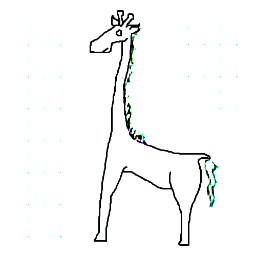}
   \end{subfigure}
   \begin{subfigure}[b]{0.08\linewidth}
      \includegraphics[width=\linewidth]{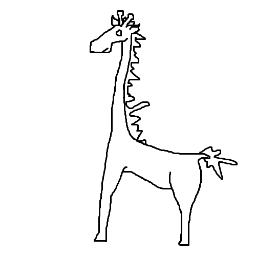}
   \end{subfigure}
   \caption{Giraffe manes: hand drawn A, AB model drawn, hand drawn target B}
   \label{fig:ABmanes}
\end{figure}

\begin{figure}
   \centering
   \begin{subfigure}[b]{0.08\linewidth}
      \includegraphics[width=\linewidth]{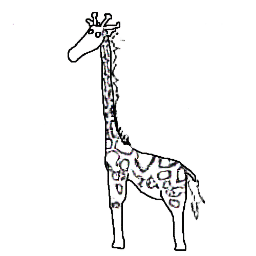}
   \end{subfigure}
   \begin{subfigure}[b]{0.08\linewidth}
      \includegraphics[width=\linewidth]{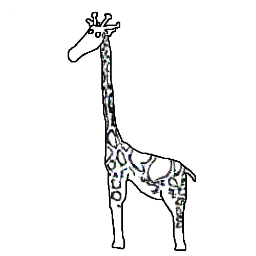}
   \end{subfigure}
   \begin{subfigure}[b]{0.08\linewidth}
      \includegraphics[width=\linewidth]{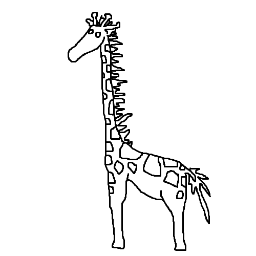}
   \end{subfigure}
   \begin{subfigure}[b]{0.08\linewidth}
      \includegraphics[width=\linewidth]{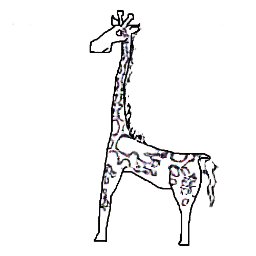}
   \end{subfigure}
   \begin{subfigure}[b]{0.08\linewidth}
      \includegraphics[width=\linewidth]{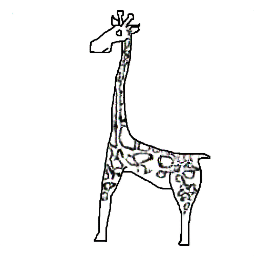}
   \end{subfigure}
   \begin{subfigure}[b]{0.08\linewidth}
      \includegraphics[width=\linewidth]{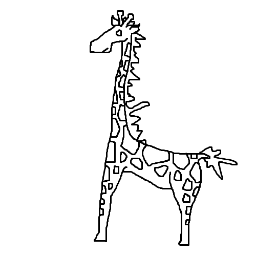}
   \end{subfigure}
   \caption{Giraffe chain test: AB + BC models, AC model, hand drawn target C}
   \label{fig: giraffechaintest}
\end{figure}

\textit{Results: }Our AB trained model, given a new body, draws that body with a good mane and tail (Fig. \ref{fig:ABmanes}). Our BC trained model, given a new body with mane and tail adds good spots. If we chain the AB and BC models together, from a new body we get good mane, tail and spots. Our AC trained model, as in the 20Dragons experiment, is worse than the two step procedure. It draws good spots but no mane or tail at all (Fig. \ref{fig: giraffechaintest}).

\section{Design Industry Applications}

Expressive augmentation and GAN drawing and coloring models, can be used as intermediate tools in the design process, creating useful variation in families of style consistent drawings, generating new ideas and allowing rapid exploration of the design space.

\subsection{Infinite Wallpaper and Scalability}

\begin{figure}
   \centering
   \begin{subfigure}[b]{0.5\linewidth}
      \includegraphics[width=\linewidth]{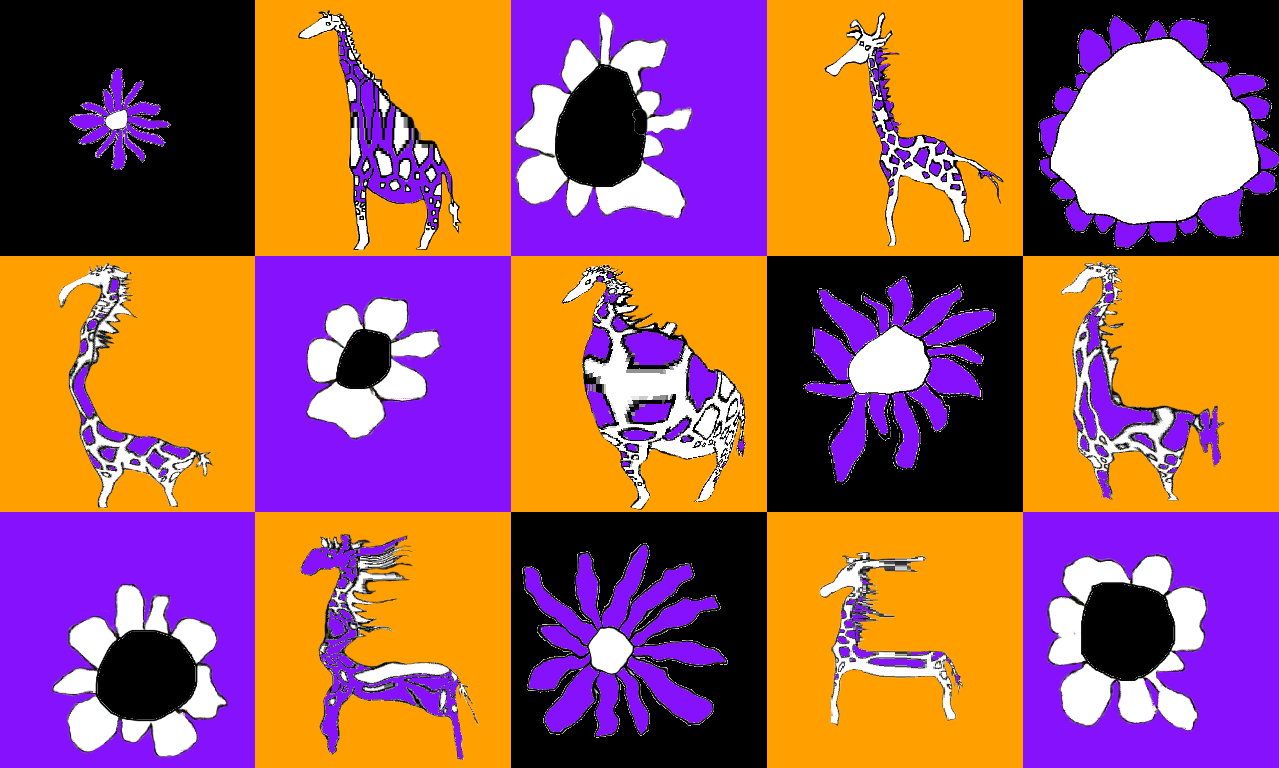}
   \end{subfigure}
   \caption{``Infinite wallpaper'' sketch}
   \label{fig:wallpaper}
\end{figure}

Fig. \ref{fig:wallpaper} is a sketch of part of an ``infinite wallpaper'' design, showing how variations of characters could be used to create regular nonrepeating patterns for wallpaper, textiles or architectural elements. In this case, the giraffe drawings and the white centered flowers were selected from the best of the expressive augmentations in our training sets. The black centered flowers have hand drawn centers and model drawn petals. We could have used giraffes with model drawn manes and spots. The coloring was done in Photoshop using the paintbrush and paint bucket tools but to scale up, the flowers and backgrounds and a modified version of the spot coloring could be done with geometric rules or GAN models exactly as we did in \citet{greene:18}.

Currently, the designer chooses the best of the augmentation drawings and model generated drawings and uses them as the foundation of finished works since the image quality is too low to use them as ready to print digital files. But our results suggest the possibility of fully automated infinite variation design families, allowing unique to the customer variation, nonrepeating regularity and "infinite wallpaper" patterns at mass market prices for any manufacturing production process where the tools can easily and cheaply change designs, including digital printers, 3D printers, laser cutters, CNC plasma cutters, CNC mills and CNC looms.     

\subsection{Augmentation and the Edge of Style}
\begin{figure}
   \centering
   \begin{subfigure}[b]{0.08\linewidth}
      \includegraphics[width=\linewidth]{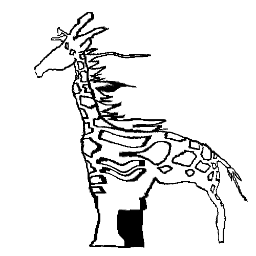}
   \end{subfigure}
   \begin{subfigure}[b]{0.08\linewidth}
      \includegraphics[width=\linewidth]{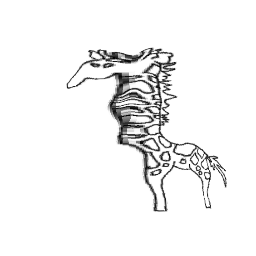}
   \end{subfigure}
   \begin{subfigure}[b]{0.08\linewidth}
      \includegraphics[width=\linewidth]{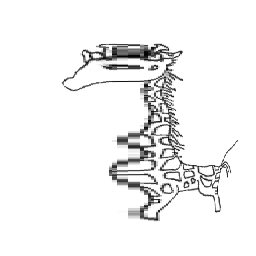}
   \end{subfigure}
   \centering
   \begin{subfigure}[b]{0.08\linewidth}
      \includegraphics[width=\linewidth]{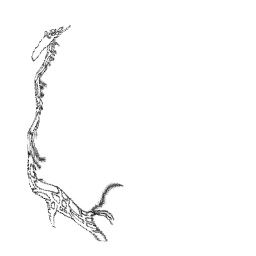}
   \end{subfigure}
   \centering
   \begin{subfigure}[b]{0.08\linewidth}
      \includegraphics[width=\linewidth]{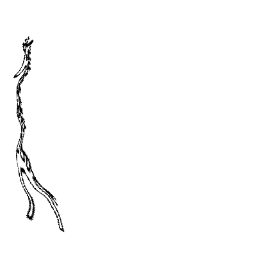}
   \end{subfigure}
   \centering
   \caption{At the edge of unacceptable: recognizable but undesirable}
   \label{fig: unacceptable}
\end{figure}

Looking for domain appropriate augmentations expanded the artist's style and increased domain knowledge about what range and kind of variation is expected, what variation is represented in our original data set, and what variation would be acceptable or desirable. The best of the expressive augmentations were chosen by the artist as more interesting than the original drawings. The more extreme augmentations, deemed unacceptable, teach us about the limits of what we're looking for (Fig. \ref{fig: unacceptable}).

\section{Conclusion and Future Work}

We have only begun to explore the space of interesting character preserving, style enhancing homeomorphisms and deformations for augmentation and image generation for drawings. In future work, we will experiment with model architecture choices and InfoGANs \citep{chen:16} and apply learned diffeomorphisms to drawings with the goal of offering the user greater variation and better control over the distribution and types of variation, enabling intuitive requests like, ``curly tails", ``fatter" or ``more novelty".

\section{Appendix: Original Drawings}

\vskip 0.2in
\bibliography{greenebibLondon}

\end{document}